\documentclass[aps,pra,twocolumn,superscriptaddress,showpacs,eqsecnum,
nofootinbib]{revtex4}

\usepackage[latin1]{inputenc}
  \usepackage[T1]{fontenc}
\usepackage{makeidx}
\usepackage[greek,francais,english]{babel}

\usepackage[dvipsone]{color}
\usepackage[dvips]{graphicx}
\usepackage{amsmath}
\usepackage{amsthm}

\newtheorem{thm}{\bf Theorem}[section]

\newtheorem{lem}[thm]{\bf Lemma}

\usepackage{latexsym}

\usepackage{amsfonts,amssymb}

\usepackage{enumerate}

\DeclareMathOperator{\Tr}{Tr}
\newcommand{\vl}{\vec{\lambda}}
\newcommand{\ml}{\mathcal{M}(\vec{\lambda})}
\newcommand{\dl}{\mathcal{D}(\vec{\lambda})}
\newcommand{\cl}{c(\vec{\lambda})}
\newcommand{\la}{\lambda}
\newcommand{\Pe}{\mathcal{P}_N}
\newcommand{\Pnn}{\mathcal{P}_{N+1}}
\newcommand{\Cd}{\mathbb{C}^d}

\newcommand{\Sl}{\mathcal{S}(\vl')}
\newcommand{\Sn}{\mathcal{S}_{N+1}}
\newcommand{\dd}{\mathrm{d}}

\begin{document}

\title{Fast rate estimation of a unitary operation in SU(d)}
\author{Jonas Kahn}
\affiliation{Universit\' e Paris-Sud 11, D\' epartement de Math\' ematiques,
B\^{a}t 425, 91405 Orsay Cedex, France}

\begin{abstract}
  We give an explicit procedure based on entangled input states for estimating
a $SU(d)$ operation $U$ with 
rate of convergence $1/N^2$ when sending $N$ particles through the device. We
prove that this rate is optimal. We also evaluate the constant $C$ such that the asymptotic risk is $C/N^2$. However other strategies might yield a better constant $C$.
\end{abstract}
\maketitle

\section{Introduction}
\label{introduction}

The question that we are investigating in this paper is: ``What is the best way
of
estimating a unitary operation $U$?''

By ``unitary operation'', we mean a device (or a \emph{channel}) that
sends a density operator
$\rho_0$ on $ \mathbb{C}^d$ to another density operator $\rho = U\rho_0 U^*$,
where $U\in SU(d)$, a special unitary matrix.

We immediately stress that the solution to this estimation problem can be
divided into two
parts: what is the input state, and which measurement (POVM) to apply on the
output state? Indeed, in order to estimate the channel $U$, we have
to let it 
act on a state (the input state). And once we have the output state,
the
problem consists in discriminating states in the family of possible output
states.

This estimation of unitary operation has been extensively studied over the last
few years. 

The first invitation
was \cite{Childs}, featuring numerous special cases. In most of those, the unitary $U$ is known to belong to some subset of
$SU(2)$. 

Then \cite{Acin} provided the form of an optimal state to be sent in
with non-specified coefficients depending on the cost function (we give the formula of this state in equation \eqref{state}).
 In that paper the authors consider the situation  where
the unitary operation is performed independently on $N $ systems. 
That study
applied to  any $SU(d)$, and any covariant loss function, in particular
fidelity, in a Bayesian framework.
The proposed input state uses an ancilla,
that is an auxiliary system that is not sent through the unitary channel with
Hilbert space $ (\mathbb{C}^d)^{\otimes N}$. The state  is prepared as a
superposition of maximally entangled states, one for each
irreducible representation of $SU(d)$ appearing in $ ( \mathbb{C}^d)^{\otimes
n}$. We emphasize that the state is an
entangled state of $( \mathbb{C}^d)^{\otimes N}\otimes ( \mathbb{C}^d)^{\otimes
N}$: we do not send $N$ copies of an entangled state through the
device, but all the $N$ systems that are sent through the channel together with
the
$N$ particles of the ancilla are part of the same entangled state, yielding
the most general possible strategy. There was no evaluation of the rate of convergence, though. 

Subsequent works mainly focused on $SU(2)$, as the case
is simpler and yields many applications, e.g.
transmission of reference frames in quantum communication. Indeed, the latter
is
equivalent
to the estimation of a $SU(2)$ operation. The first strategy to be proved to
converge (in
fidelity) at $1/N^2$ rate was not covariant \cite{Peres}. It made no use of an
ancilla. Later, the same rate was achieved for a covariant measurement with an
ancilla \cite{bag} through a judicious choice of the coefficients left free in the state
proposed in \cite{Acin}. The optimal constant ($\pi^2/N^2$ for
the fidelity) was also
computed. It was almost simultaneously noticed
\cite{bag2, chiribella} that asymptotically the ancilla is unnecessary. Indeed
what we need is entangling different copies of the same irreducible
representation. Now each irreducible representation appears with 
multiplicity in $( \mathbb{C}^d)^{\otimes N}$, most of them with higher multiplicity
than dimension, which is the condition we need. This method was dubbed
``self-entanglement''. The advantage is that we need to prepare half the  number of  
particles, as we do not need an ancilla. In all these articles,  the Bayesian paradigm with
uniform prior was used. The same $1/N^2$ rate was shown to hold true in a minimax sense,
in pointwise estimation
\cite{hay}. We stress the importance of this $1/N^2$ rate, proving how
useful entanglement can be. Indeed, in classical data analysis, we cannot
expect a better rate than $1/N$. Similarly the  $1/N$ bound holds  for any
strategy where
the $N$ particles we send through the device are not entangled ``among
themselves'' (that is, even if there is an ancilla for each of these $N$
particles).

Another popular theme has been the determination of the phase $\phi$ for unitaries of the form $U_{\phi} = e^{i\phi H}$. This very special case already has many applications, especially in interferometry or measurement of small forces, as featured in the review article \cite{review} and references therein. A common feature of the most efficient techniques is the need for entangled states of many particles, and much experimental work has aimed at generating such states. These methods essentially involve either manipulation of photons obtained through parametric down-conversion (for example \cite{eisenberg}), ions in ion traps (for example \cite{ions}) or atoms in cavity QED (for example \cite{raimond}).

In recent years, there has been renewed interest in the $SU(d)$ case. Notably,
\cite{chiribella2} takes off from \cite{Acin}, allowing for more general
symmetries and making explicit for natural cost functions both the free coefficients -- as the coordinates of
the
eigenvector of a matrix -- and the POVM (see
Theorem \ref{chiri} below).  
With a completely different strategy,
aiming rather at pointwise estimation (and therefore minimax theorems), an
input state for $U^{\otimes n}$ was found \cite{bal, thesisball} such that the Quantum Fisher
Information matrix is scaling like $1/N^2$, yielding hopes of getting as
fast an estimator for $SU(d)$. No associated measurement was found in that
paper.

Given the state of the art, a natural question is whether we can obtain, as for
$SU(2)$, this
dramatic increase in performance when using entanglement for general
$SU(d)$. That is, do we have an estimation procedure whose rate
is $1/N^2$, instead of $1/N$?  Neither \cite{chiribella2}, where  the
asymptotics are not studied for $SU(d)$, nor \cite{bal},
where no measurement is given, answer
this question.  

\medskip

In this article, we first prove that we cannot expect a better rate than
$1/N^2$. This kind of bound based on the laws of quantum physics, without any \emph{a priori} on the experimental device, is traditionally called the \emph{Heisenberg limit} of the problem. Then we choose a completely explicit input state of the form \eqref{state} (as in \cite{Acin}), by specifying  the coefficients. By using the associated POVM, the estimator of
a unitary quantum operation $U\in SU(d)$ converges   at rate $1/N^2$. The
constant is not optimal, but is briefly studied at the end of the paper. We obtain these results with fidelity as a cost function, both in a Bayesian setting, with a uniform prior, and in a minimax setting. Notice that we shall not need
an ancilla.

The next section consists in formulating the problem and
restating Theorem 2 of \cite{chiribella2} within our framework. Section \ref{uneff} then
shows that it is impossible to converge at rate faster than $O(N^{-2})$. In
section \ref{cost}, we write a general formula for the risk of a strategy as
described in Theorem \ref{chiri}, and in section \ref{choice} we specify our
estimators by choosing our coefficients in (\ref{state}). We then prove that
the risk of this estimator is $O(N^{-2})$. The last section
(\ref{constante}) consists in finding
the precise asymptotic speed of our procedure, that is the constant $C$ in $C
N^{-2}$. We finish by stating in Theorem \ref{conclusion} the results of the paper.

\section{Description of the problem}
\label{notations}

We are given an unknown unitary operation $U\in SU(d)$ and must estimate it ``as
precisely as possible''. We are allowed to let it act on $N$ particles, so
that we are discriminating between the possible $U^{\otimes N}$. We shall work both with pointwise estimation (as
preferred by mathematicians) and with a Bayes uniform prior (a favorite of
physicists).

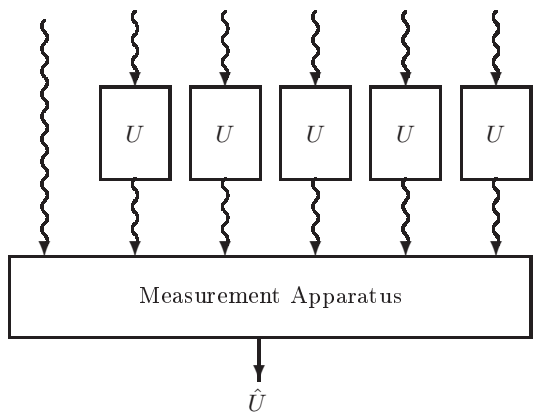
\begin{figure}[ht]
\begin{center}
\setlength{\unitlength}{0.30cm}
\begin{picture}(27,20)(0,0)
\thicklines
\multiput(6,12)(4,0){5}{\framebox(3,4){$U$}}
\thinlines
\thicklines
\put(3.5,9){
\multiput(0,0)(0,1){10}{\qbezier(0,0)(0.25,0.25)(0,0.5)\qbezier(0,0.5)(-0.25,0.75)(0,1)}
}
\multiput(7.5,16.4)(4,0){5}
{
\multiput(0,0)(0,1){3}{\qbezier(0,0)(0.25,0.25)(0,0.5)\qbezier(0,0.5)(-0.25,0.75)(0,1)}
}
\multiput(7.5,16)(4,0){5}{\vector(0,-1){0}}
\multiput(7.5,9)(4,0){5}
{
\multiput(0,0)(0,1){3}{\qbezier(0,0)(0.25,0.25)(0,0.5)\qbezier(0,0.5)(-0.25,0.75)(0,1)}
}
\multiput(3.5,8.5)(4,0){6}{\vector(0,-1){0}}
\put(2,5){\framebox(23,3.5){Measurement Apparatus}}
\put(13,5){\vector(0,-1){2}}\put(12.5,1.7){$\hat{U}$}
\end{picture}
\caption{\label{fwithentanglement} Most general estimation scheme of $U$ when $n$ copies are available at
the same time, and using entanglement.}
\end{center}
\end{figure}

Any estimation procedure can be described as follows (see Figure
\ref{fwithentanglement}):  the unitary channel $U^{\otimes N}$ acts as 
\[
U^{\otimes N} \otimes {\bf 1} : (\mathbb{C}^{d})^{\otimes N} \otimes
\mathcal{K} \to (\mathbb{C}^{d})^{\otimes N} \otimes
\mathcal{K}, 
\]
on the space of the $N$ systems together with a possible ancilla. The 
input state $\rho_n\in M(
(\mathbb{C}^{d})^{\otimes n} \otimes \mathcal{K}_n)$ is mapped into an output
state on which we perform a measurement $M$  whose result is the estimator $\hat{U}\in SU(d)$.

In order to evaluate the quality of an estimator $\hat{U}$, we fix a cost
function $\Delta(U,V)$. The
global pointwise risk of the estimator is
\begin{equation*}
R_{P}(\hat{U}) = \sup_{U\in SU(d)} \mathbb{E}_U[\Delta(U,\hat{U})].
\end{equation*} 
The probability distribution of $\hat{U}$ depends on $U$, and we take expectation
with respect to this probability distribution.

On the other hand, the Bayes risk with uniform prior is:
\begin{equation*}
R_{B}(\hat{U}) = \int_{SU(d)} \mathbb{E}_U[\Delta(U,\hat{U})] \dd \mu(U).
\end{equation*} 
where $\mu$ is the Haar measure on $SU(d)$.

As cost function, we choose the fidelity $F$ (or rather $1-F$), which
for an element of $SU(d)$ is defined as:
\begin{align*}
\Delta(U,\hat{U}) & = 1 - \frac{|\Tr(U^{-1}\hat{U})|^2}{d^2} \\
                  & = 1 - \frac{|\chi_{\Box}(U^{-1}\hat{U})|^2}{d^2}
\end{align*} 
where $\chi_{\Box}$ is the character of the defining representation of $SU(d)$,
whose Young tableau consists in only one box. In other words, $\chi_{\Box}(U) =
\Tr(U)$.

\smallskip

Before really addressing the problem, we make a few remarks on why this choice
of distance is suitable for mathematical analysis.

Firstly, this cost function is covariant, i.e. $\Delta(U,\hat{U}) = \Delta({\bf
1}_{\Cd}, U^{-1}\hat{U})$.

Secondly,  a useful feature within the Bayesian framework is that $\Delta$ is
of the form (\ref{charac}), as  required in Theorem \ref{chiri}. Indeed we can
rewrite $\Delta(U,\hat{U})$ as  $1 -
\chi_{\Box}(U^{-1}\hat{U})\chi_{\Box}^*(U^{-1}\hat{U})/d^2$. Now the conjugate
of a character is the character of the adjoint representation, the product of
two characters is again the character of a possibly reducible representation
$\pi$. This character is equal to the sum of the characters of the irreducible
representations appearing in the Clebsch-Gordan development of $\pi$, in which
all coefficients are non-negative. Therefore $\Delta = 1 - (\sum_{\vl} a_{\vl}
\chi_{\vl}^* )$ where $a_{\vl}\geq 0$ and $\vl$ runs over all irreducible
representations of $SU(d)$. That is the condition (\ref{charac}) that we shall need for applying Theorem \ref{chiri}, given at the end of the section. 

On the other hand, the theory of pointwise estimation deals usually with the
variance of the estimated parameters when we use a smooth parameterization of
$SU(d)$. As we want to use the Quantum Cram\'er-Rao Bound (\ref{QCRB}), we need
$\Delta$ to be quadratic in the parameters to the first order, and positive
lower bounded for $\hat{U}$ outside a neighborhood of $U$. As $\Delta$ is
covariant, it is sufficient to check this with $U={\bf 1}_{ \mathbb{C}^d}$. Now
an example of a smooth parameterization in a neighborhood of the identity is
$U(\theta) = \exp(\sum_{\alpha}\theta_\alpha T_{\alpha})$ where $\theta\in
\mathbb{R}^{d^2-1}$ and the $T_{\alpha}$ are generators of the Lie algebra, so
that $\Tr(T_{\alpha}) = 0$. Now $\Tr[ \exp(\sum_{\alpha}\theta_\alpha
T_{\alpha}) ] = d + \sum_{\alpha} \theta_{\alpha} \Tr(T_{\alpha}) +
O(\|\theta\|^2)$, so that the trace minus $d$, and consequently $\Delta$, is
quadratic in $\theta$ to the first order.

\medskip

As stated at the beginning of this section, we are working with
$U^{\otimes N}$. The Clebsch-Gordan decomposition of the $n$-th tensor product
representation is  
\[
U^{\otimes N}= \bigoplus_{\vl : |\vl|=N} U^{\vl} \otimes {\bf 1}_{
\mathbb{C}^{\ml}}
\]
acting on $\bigoplus_{\vl : |\vl|=N} \mathcal{H}^{\vl} \otimes
\mathbb{C}^{\ml}$, where $ \mathcal{H}^{\vl} = \mathbb{C}^{\dl}$ is the
representation space of $\vl$, $\ml$ is the multiplicity of $\vl$ in the
$n$-th tensor product representation, and $\dl$ the dimension of $\vl$. We
refer to $ \mathbb{C}^{\ml}$ as the multiplicity space of $\vl$. We have
indexed the irreducible representations of $SU(d)$ by $\vl = (\lambda_1,\dots,\lambda_d)$, and
written $|\vl| = \sum_{i=1}^d \lambda_i$. Notice that this labelling of
irreducible representations
is redundant, but that if $|\vec{\lambda}^1| = |\vec{\lambda}^2|$, then $\vl^1$
and $\vl^2$ are equivalent (denoted $\vl^1\equiv\vl^2$) if and only if $\vl^1=\vl^2$. 

The starting point of our argument will be the following reformulation  of the results of \cite{chiribella2}, with less
generality, and without the formula for the risk whose form is not adapted to
our subsequent analysis: 
\begin{thm}\emph{\cite{chiribella2}}
\label{chiri}
Let $U\in SU(d)$ be a unitary operation to be estimated, through its action on
$N$ particles. We may use entanglement and/or an ancilla. 

Then, for a uniform prior and any cost function of the form
\begin{equation}
\label{charac}
c(U,\hat{U}) = a_0 - \sum_{\vl} a_{\vl} \chi^*_{\vl}(U^{-1}\hat{U}),
\end{equation} 
we can find as optimal input state a pure state of the form 
\begin{equation}
\label{state}
|\Psi\rangle = \bigoplus_{\vl : |\vl| =N} \frac{\cl}{\sqrt{\dl}}
\sum_{i=1}^{\dl} |\psi^{\vl}_i\rangle\otimes|\phi^{\vl}_i\rangle
\end{equation} 
with $\cl\geq 0$, and the normalization condition,
\begin{equation}
\label{trace1}
\sum_{\vl} \cl^2 = 1.
\end{equation}
Moreover $|\psi^{\vl}_i \rangle$ is an orthonormal basis of $
\mathcal{H}^{\lambda}$ and $|\phi^{\vl}_i\rangle$ are orthonormal vectors of
the multiplicity space, which may be augmented by an ancilla if necessary (see remark below on the dimensions).

The corresponding measurement is the covariant POVM with seed $\Xi =
|\eta\rangle\langle \eta|$ given by:
\begin{equation}
\label{seed}
|\eta\rangle = \bigoplus_{\vl| \cl\neq 0} \sqrt{\dl} \sum_{i=1}^{\dl}
|\psi^{\vl}_i\rangle\otimes|\phi^{\vl}_i\rangle,
\end{equation} 
that is a POVM whose density with respect to the Haar measure is given by
$m(U)=U|\eta\rangle\langle \eta| U^*$ with
\[
U |\eta\rangle = \bigoplus_{\vl| \cl\neq 0} \sqrt{\dl} \sum_{i=1}^{\dl}
 U^{\vl} |\psi^{\vl}_i\rangle\otimes|\phi^{\vl}_i\rangle.
\]
\end{thm}

\emph{Remark:} We use $\dl$ orthonormal vectors in the multiplicity
space of $\vl$. This requires $\ml\geq \dl$. If this is not the case, we must
increase the dimension of the multiplicity space by using an ancilla in $
\mathbb{C}^\delta$. Then the action of $U$ is $U^{\otimes N}\otimes {\bf 1}_{
\mathbb{C}^{\delta}}$ whose Clebsch-Gordan decomposition is $\bigoplus_{\vl | |\vl|=N} U^{\vl} \otimes {\bf 1}_{
\mathbb{C}^{\delta\ml}}
$. With big enough $\delta$, we have $\delta\ml\geq \dl$. Notice that an
ancilla is not necessary if $\cl = 0$ for all $\vl$ such that $\dl>\ml  $.

Another remark is that, as defined, our POVM is not properly normalized: 
$M(SU(d))\neq
{\bf 1}$, but is equal to the projection on the space spanned by the
$U|\Psi\rangle$. As this is the only subspace of importance, we can complete
the POVM (through the seed, for example) \emph{ad libitum}.

\medskip

Our estimator $\hat{U}$ is the result of the measurement with POVM defined
by \eqref{seed} and input state of the form \eqref{state}, with specific $\cl$. 
Such an estimator is covariant, that is $p_U(\hat{U}) = p_{{\bf
1}_{\mathbb{C}^d}}(U^{-1}\hat{U})$, where $p_U$ is the probability distribution
of $\hat{U}$ when we
are estimating $U$. The cost function is also covariant, so
that $ \mathbb{E}_{U}[\Delta(U, \hat{U})] $ does not depend on $U$. This
implies that the Bayesian risk and the pointwise risk coincide. With the
second equality true for all $U\in SU(d)$, we have:
\begin{equation}
\label{equiv}
R_B(\hat{U}) = R_P(\hat{U}) = \mathbb{E}_{U}[\Delta(U, \hat{U})]  .
\end{equation}

Theorem \ref{chiri} states that there exists an optimal (Bayes uniform)
estimator $\hat{U}_o$ of this form (corresponding to the optimal choice of
$\cl$), so that it obeys
(\ref{equiv}). From this we first prove that no estimator whatsoever can have a
better rate than $1/N^2$.

\section{Why we cannot expect better rate than $1/N^2$}
\label{uneff}

For proving this result, we need the Bayesian risk for priors $\pi$ other than
the uniform prior:
\[
R_{\pi}(\hat{U}) = \mathbb{E}_{\pi}[\mathbb{E}_U[\Delta(U,\hat{U})]].
\]

As $\hat{U}_o$ is Bayesian optimal for the uniform prior, we only have to prove
that $R_B(\hat{U}_o)=
O(N^{-2})$. This is also sufficient for pointwise risk as, for any estimator
$\hat{U}$, we have $R_B(\hat{U})\leq R_P(\hat{U})$.
Moreover, as  $ \mathbb{E}_{U}[\Delta(U, \hat{U}_o)] $ does not depend on $U$,
$R_{\pi}(\hat{U}_o) = R_B(\hat{U}_o)$. It is then sufficient to prove, for a
$\pi$ of our choice, that:
\begin{equation}
\label{objectif}
R_{\pi}(\hat{U}_o) = O(N^{-2}).
\end{equation} 

The idea is to find a Cram\'er-Rao bound that we can apply to some $\pi$. We
shall combine the Braunstein and Caves information inequality (\ref{BCII}) and
the Van Trees inequality (\ref{VanTree}) to obtain the desired Quantum
Cramér-Rao Bound, much in the spirit of \cite{gill1}. This bound will yield an
explicit rate through a result of \cite{bal}.

Van Trees' inequality states that given a classical statistical model smoothly
parameterized by $\theta\in \Theta\subset \mathbb{R}^p$, and a smooth
prior with compact support $\Theta_0\subset \Theta$, then for any estimator
$\hat{\theta}$, we have:
\begin{equation}
\label{VanTree}
\mathbb{E}_{\pi}[\Tr(V_{{\theta}}(\hat{\theta}))] \geq \frac{ p^2}{
\mathbb{E}_{\pi}[\Tr(I(\theta))] -
\mathcal{I}_{\pi}},
\end{equation} 
where $I(\theta)$ is the Fisher information matrix of the model at point
$\theta$, $ \mathcal{I}_{\pi}$ is 
a finite (for reasonable $\pi$) constant depending on $\pi$ (quantifying in
some way the prior information), and $V_{\theta}(\hat{\theta})\in M_p( \mathbb{R})$ is the mean square error (MSE) of the
estimator $\hat{\theta}$ at point $\theta$ given by: 
\[
V_{{\theta}}(\hat{\theta})_{\alpha,\beta} =  \mathbb{E}[(\theta_{\alpha} - \hat{\theta}_{\alpha})(\theta_{\beta}
-\hat{\theta}_{\beta})]. 
\]
This form of Van Trees inequality is obtained by setting $N=1$, $G= C= Id$ and $\psi
= \theta$ in (12) of \cite{gill1}.

Now the Braunstein and Caves information inequality \cite{Braunstein&Caves} yields an upper bound on the information matrix  $I_M(\theta)$ of any
classical 
statistical model obtained by applying the measurement $M$ to a quantum
statistical model.
For any family of quantum states parameterized  by a $p$-dimensional parameter
$\theta\in \Theta\in \mathbb{R}^p$, for any measurement $M$ on these states,
the following holds:
\begin{equation}
\label{BCII}
I_M(\theta)\leq H(\theta),
\end{equation} 
where $H(\theta)$ is the quantum Fisher information information matrix at point $\theta$.

Now it was proved in \cite{bal} that for a smooth parameterization of an open
set of $SU(d)$, and for any input state, the quantum Fisher information of the
output states fulfils:
\[
H(\theta) = O(N^2).
\] 
Inserting in (\ref{VanTree}) together with (\ref{BCII}) we get as quantum Cram\'er-Rao bound
\begin{equation}
\label{QCRB}
\mathbb{E}_{\pi}[\Tr(V_{{\theta}}(\hat{\theta}))] = O\left(\frac1{N^{2}}\right).
\end{equation}

\smallskip

We now want to apply this bound to  obtain \eqref{objectif}. There are a few small technical difficulties. 
First of all, we cannot use the 
uniform prior for $\pi$ as  $SU(d)$ is not homeomorphic to an open set of
$\mathbb{R}^p$. We then have to define two neighborhoods
of the identity $\Theta_0 \subset \Theta$, allowing to use the Van Trees inequality. Now our
estimator $\hat{U}_o$ need not be in $\Theta$, so that we shall in fact
apply Van Trees inequality to a modified estimator $\tilde{U}$. Finally, this
bound is on the variance, and we must relate it to $\Delta$.

 Our first task consists
in restricting our attention to a neighborhood $\Theta$ of ${\bf 1}_{
\mathbb{C}^d}$. It corresponds to a neighborhood $\Theta$ (we use
the same notation) of $0\in
\mathbb{R}^p$ through $U=\exp(\sum_{\alpha}\theta_{\alpha}T_{\alpha})$. This holds if the neighborhood is small enough, so we
define it by $U\in\Theta$ if and only if $\Delta({\bf 1}_{
\mathbb{C}^d}, U) < \epsilon$ for a fixed small enough $\epsilon$. We
define $\Theta_0$ through $U\in \Theta_0$ for  $\Delta({\bf 1}_{
\mathbb{C}^d}, U) \leq \epsilon/3$, and take a smooth fixed prior $\pi$
with support in $\Theta_0$, such that $ \mathcal{I}_{\pi}<\infty$.

Now we modify our estimator $\hat{U}_o$ into an estimator $\tilde{U}$ given by
$\tilde{U}=\hat{U}_o$ for $\hat{U}_o\in \Theta$ and $\tilde{U} = {\bf
1}_{\mathbb{C}^d}$ for $\hat{U}_o\not\in\Theta$. Then, by the triangle
inequality,  for any $U\in\Theta_0$, we
have $\Delta(U,\hat{U}_o)\geq \Delta(U,\tilde{U})$. 

The fundamental point of the reasoning (used at (\ref{used})) is that, as $\Delta$ is quadratic at the
first-order, there is a positive constant $c$ such that, for any $U^1,U^2\in
\Theta$, corresponding  to $\theta^1, \theta^2$, we have $\Delta(U_1, U_2)\geq
c \sum_{\alpha} (\theta^1_{\alpha}-\theta^2_{\alpha})^2$.

Finally we get
\begin{align}
R_{\pi}(\hat{U}_o) & = \mathbb{E}_{\pi}[\mathbb{E}_{U}[\Delta(U,\hat{U}_o)]] \notag\\
& \geq \mathbb{E}_{\pi}[\mathbb{E}_{U}[\Delta(U,\tilde{U})]] \notag\\
\label{used}
& \geq c \mathbb{E}_{\pi}[V_{\tilde{\theta}}] \\
& = O(N^{-2}). \notag
\end{align}

We have thus proved (\ref{objectif}), and hence our bound on the efficiency of any estimator.

We now write formulas for the risk of any estimator of the form given in
Theorem \ref{chiri}.

\section{Formulas for the risk}
\label{cost}

By (\ref{equiv}), our risk $R_P(\hat{U})$ is equal to the pointwise risk at
${\bf
1}_{\mathbb{C}^d}$, with which we shall work:
\begin{align}
\label{etape1}
 \int_{SU(d)} p_{{\bf
1}_{ \mathbb{C}^d}}(\hat{U}) 
\left\{1 - \frac{|\chi_{\Box}(\hat{U})|^2}{d^2}\right\} d\mu(\hat{U}).  
\end{align} 

Now we compute the probability distribution  of $\hat{U}$ for a given
$|\Psi\rangle$ of the form \eqref{state}, that is 
\begin{align*}
p_{{\bf
1}_{ \mathbb{C}^d}}(\hat{U}) &= \langle \Psi | \hat{U} \Xi \hat{U}^* | \Psi\rangle \\
& = \left| \sum_{\vl : |\vl|=N} \frac{\cl}{\dl} \dl \sum_{i=1}^{\dl} \langle
\psi^{\vl}_i | U | \psi^{\vl}_i\rangle \right|^2 \\
& = \left|\sum_{\vl : |\vl|=N} \cl \chi_{\vl}(\hat{U})\right|^2,
\end{align*}
where we have used that the character $\chi_{\vl}$ of $\vl$ is the trace of
$U$ in the representation.

Then, using (\ref{etape1}), recalling that $p_{{\bf 1}_{\mathbb{C}^d}}$ is a
probability density for Haar measure $\mu$ on $SU(d)$,    and that
$\chi_{\vl^1}\chi_{\vl^2} = 
\chi_{\vl^1\otimes\vl^2}$ (for the second term), we get: 
\begin{equation}
\begin{split}
 R_P(\hat{U})  
 = 1 - 
\frac1{d^2}\int_{SU(d)} \left|\sum_{\vl :
|\vl|=N}\cl\chi_{\vl\otimes\Box}(\hat{U})\right|^2 d\mu(\hat{U}).
\label{general}
\end{split}
\end{equation}

In order to evaluate the second term, we use  the following orthogonality relations for characters: 
\begin{equation}
\label{orth}
\int_{SU(d)}d\mu(U) \chi_{\vl_1}(U)
\chi_{\vl_2}(U)^* =
\delta_{\vl_1\equiv\vl_2}. 
\end{equation} 

To do so we need the Clebsch-Gordan series of $\vl\otimes \Box$:
\begin{equation}
\label{CG}
\vl \otimes \Box = \oplus_{\{1\leq i\leq d | \lambda_i>\lambda_{i+1} \}} \vl +
e_i,
\end{equation} 
where conventionally $\lambda_{d+1}=0$. Here we see $\vl$ as a $d$-dimensional
vector and $e_i$ as the $i$-th basis vector.

We then reorganize the sum of characters as:
\begin{align*}
\sum_{\vl : |\vl|=N}\cl \chi_{\vl\otimes\Box}(\hat{U}) =
\sum_{\vl' : |\vl'|=N+1}\sum_{i\in\mathcal{S}(\vl')}
c(\vl'-e_i)\chi_{\vl'}(\hat{U}),
\end{align*} 
where $\Sl$ is the set of $i$ between $1$ and $d$ such that $\vl'-e_i$ is still
a representation, that is  $\la'_i>\la'_{i+1}$. We shall write $\#\Sl$ for its
cardinality. 

Inserting in (\ref{general}) and remembering (\ref{orth}), we are left with 
\begin{equation}
\label{espdelta}
R_P(\hat{U}) = 1 - \frac{\sum_{\vl' :
|\vl'|=N+1} |\sum_{i\in\Sl} c(\vl'-e_i)|^2}{d^2}.
\end{equation} 

To go any further, we must work with specific $\cl$.

\section{Choice of the coefficients $\cl$ and proof of their efficiency}
\label{choice}

We now have to choose the coefficients $\cl$ so that the
right-hand side of (\ref{espdelta}) is small.

It appears useful to introduce  subsets of the set of all irreducible
representations. Let $\Pe = \{\vl |\;|\vl|=N;
\la_1>\dots>\la_d>0\}$. Obviously, if $\vl'\in\Pnn$, then $\#\Sl=d$, and the
converse is true. We can see them intuitively as points on a
$(d-1)$-dimensional surface, and with this picture in mind, we shall speak of
the border of $\Pe$ (when $\la_i = \la_{i+1} +1$ for some $i$), or of being far
from the border (without precise
mathematical meaning).

We are ready to give heuristic arguments on how good coefficients should behave.

We must try to get the fraction in (\ref{espdelta}) close to one. Now 
\begin{align*}
&\frac{\sum_{\vl' :
|\vl'|=N+1} |\sum_{i\in\Sl} c(\vl'-e_i)|^2}{d^2} \\ 
& \leq \sum_{\vl' :
|\vl'|=N+1}  \frac{\#\Sl}{d} \frac{\sum_{i\in\Sl} |c(\vl'-e_i)|^2}{d} \\
& \leq  \sum_{\vl' :
|\vl'|=N+1}   \frac{\sum_{i\in\Sl} |c(\vl'-e_i)|^2}{d}
\\
& \leq \sum_{\vl : |\vl| = N} |\cl|^2 = 1.
\end{align*}
The first inequality was obtained using Cauchy-Schwarz inequality for each inner sum.
There is equality if $c(\vl'-e_i)$ does not depend on $i$. From this, we deduce
that for most $\vl'$, the $c(\vl'-e_i)$ must be approximately equal, especially
if they are large. The second inequality follows from $\#\Sl\leq d$. From
this we deduce that for $\vl\not\in\Pnn$, the coefficients $c(\vl-e_i)$ must be
small. Remark that about $1/N$ of the $\vl'$ such that $|\vl'|=N+1$ are not in
$\Pnn$, so that if all $\cl$ were equal, these border terms would cause our rate
to be $1/N$. The key of the third inequality is to notice that each $\cl$ is
appearing in the sum once for each term in its Clebsch-Gordan series
(\ref{CG}), and that there are at most $d$ terms. Please note that there are
$d$ terms if $\vl\in\Pe$, and  if $\vl'$ is in $\Pnn$, far from the border,
then $\vl'-e_i$ is in $\Pe$, far from the border.

The conclusion of these heuristics is that we must choose coefficients
``locally'' approximately equal (at most $1/N$ variation in ratio), and that
the coefficients must go to $0$ when we are approaching the border of $\Pe$.

\medskip

One weight satisfying these heuristics is the following.
\begin{equation}
\label{weight}
\cl = \mathcal{N}\prod_{i=1}^d p_i,
\end{equation} 
where $\mathcal{N}$ is a normalization constant to ensure that (\ref{trace1})
is satisfied and $p_i=\la_i-\la_{i+1}$. 
 We shall use it below,
and prove that it delivers the $1/N^2$ rate.

A first remark about these weights is that $\cl = 0$ if $\vl\not\in\Pe$. Now,
for any $\vl\in\Pe$, we have $\dl\geq \ml$, so that we do not need an ancilla.

Indeed, using hook formulas (see \cite{Schensted}), we get $\ml/\dl=
N!\prod_{i=1}^d\frac{(\lambda_i+d-i)!}{(d-i)!}$. Now for $\vl\in \Pe$, we
know that $\la_i\neq 0$. Under this constraint and $\sum \la_i =N$, the
maximum is attained by $\lambda_1 =N-d+1$ and $\lambda_i =1$ for $i\neq1$. We
end up with exactly $1 $.

\smallskip

We shall now use (\ref{weight}) and express the numerator of (\ref{espdelta})
with our choice of $p_i$. Notice
first that if $p_j$ characterize $\vl'$ then those which characterize $\vl'-e_i$
are given by $p_j^{(i)} = p_j + \delta_{j,i-1} - \delta_{j,i}$. So 
\begin{equation*}
 \mathcal{N}^{-1} c(\vl'-e_i) = \prod_{j=1}^d p_j + r_{\vl'}(i),
\end{equation*} 
with
\begin{equation*}
r_{\vl'}(i) = - \prod_{j\neq i} p_j + \delta_{j>1}\left(\prod_{j\neq i-1}
p_j -\prod_{j\neq i, i-1} p_j\right).
\end{equation*} 
Introducing another notation will make this slightly more compact. For a vector
$\vec{x}$ with $d$ components and $\mathcal{E}$ a subset of $\{1,\ldots,d\}$, define:
\begin{equation}
\label{notation}
x_{\mathcal{E}} = \prod_{j\neq \mathcal{E}} x_j.
\end{equation}
Then 
\[
r_{\vl'}(i) = - p_{\{i\}} + \delta_{j>1}\left(p_{\{i-1\}}
-p_{\{i, i-1\}}\right).
\]

Notice now that for $\vl\in \Pe$, there are exactly $d$ irreducible
representations appearing in
the Clebsch-Gordan decomposition of $\vl\otimes\Box$ (\ref{CG}). So that
$c(\vl)^2$
appears exactly $d$ times in $\sum_{\vl' : |\vl'|=N+1}\sum_{i\in\Sl}
c(\vl'-e_i)^2$.
We may then rewrite the renormalization constant $\mathcal{N}$ as
\[
d^{-1}  \sum_{\vl' :
|\vl'|=N+1} \sum_{i\in\Sl} \prod_{j=1}^d p_j^{(i)2}.
\]

Therefore, rewriting the second term in (\ref{espdelta}) with our values of
$\cl$, we aim at proving:
\begin{equation}
\label{seek}
 \frac{\sum_{\vl' :
|\vl'|=N+1} \left(\sum_{i\in\Sl} \prod_{j=1}^d p_j + r_{\vl'}(i)\right)^2}
{d  \sum_{\vl' :
|\vl'|=N+1} \sum_{i\in\Sl} \left(\prod_{j=1}^d p_j + r_{\vl'}(i)\right)^2 }
= 1 +O(N^{-2}).
\end{equation}

Let us expand the numerator:
\begin{align*}
\sum_{\vl' :
|\vl'|=N+1} \left(\sum_{i\in\Sl} \prod_{j=1}^d p_j + r_{\vl'}(i)\right)^2 
=
C_t\left(1 + t_1 + t_2\right),
\end{align*} 
with
\begin{align*}
C_t & = \sum_{\vl'} (\#\Sl)^2  \prod_{j=1}^d p_j^2,
 \\
t_1 & =\frac{2 \sum_{\vl'}\sum_{i\in \Sl}\#\Sl r_{\vl'}(i)\prod_{j=1}^d
p_j}{ C_t},
 \\
t_2 & =\frac{\sum_{\vl'}\left(\sum_{i\in\Sl} r_{\vl'}(i)\right)^2}{C_t}.
\end{align*}

Similarly the denominator can be read as:

\begin{align*}
d \sum_{\vl' :
|\vl'|=N+1} \sum_{i\in\Sl} \left(\prod_{j=1}^d p_j + r_{\vl'}(i)\right)^2 
=
C_u \left(1 + u_1+u_2  
\right),
\end{align*}
with
\begin{align*}
C_u & = \sum_{\vl'} d\#\Sl  \prod_{j=1}^d p_j^2, 
\\
u_1 & = \frac{2d \sum_{\vl'}\sum_{i\in \Sl}
r_{\vl'}(i)\prod_{j=1}^d
p_j}{ C_u},
\\
u_2 & =\frac{\sum_{\vl'} d\sum_{i\in\Sl} r_{\vl'}(i)^2}{C_u}.
\end{align*}

With these notations, we aim at proving the set of estimates given in Lemma
\ref{orders}.
Indeed they imply:
\begin{equation}
\begin{split}
\label{result}
 \frac{\sum_{\vl' :
|\vl'|=N+1} \left(\sum_{i\in\Sl} \prod_{j=1}^d p_j + r_{\vl'}(i)\right)^2}
{d  \sum_{\vl' :
|\vl'|=N+1} \sum_{i\in\Sl} \left(\prod_{j=1}^d p_j + r_{\vl'}(i)\right)^2 }\\
= 1 + t_2 - u_2 +  O(N^{-3})
\end{split}
\end{equation}  
with $(t_2-u_2)$ of order $N^{-2}$. By (\ref{seek}), the risk of the estimator
is then $u_2-t_2 + O(N^{-3})$. Thus proving Lemma \ref{orders} amounts at
proving $1/N^2$ rate.


We shall make use of the notation $\Theta(f)$, meaning that there are universal
positive constants $m$ and $M$ such that:
\[
m f \leq \Theta(f) \leq M f.
\]

\begin{lem}
\label{orders}
With the above notations, 
\begin{align*}
C_u & = C_t =  d^2 \sum_{\vl' :
|\vl'|=N+1}  \left(\prod_{j=1}^d p_j \right)^2 \\
& = \Theta(N^{3d-1}) \\
t_1& = u_1 = O(N^{-1}) \\
t_2 &= O(N^{-2})\\
u_2 & = O(N^{-2}).
\end{align*}
\end{lem}

\begin{proof}
We first prove the first line.

Indeed for $\vl'\in\Pnn$, all $i$ are in $\Sl$, and $\left(\sum_{i\in\Sl}
\prod_{j=1}^d p_j \right)^2 = d \sum_{i\in\Sl}
\prod_{j=1}^d p_j^2 =d^2 \prod_{j=1}^d p_j^2$. But if $\vl'\not\in\Pnn$,
there is at least one $p_j$ equal to zero, so they do not contribute to the
sum. So that $C_u = C_t =  d^2 \sum_{\vl' :
|\vl'|=N+1}  \left(\prod_{j=1}^d p_j \right)^2$. 

We have then equality of the denominators of $t_1$ and $u_1$. The same argument
gives equality of the numerators. On $\Pnn$, $\#\Sl = d$ so that 
\[
\sum_{i\in
\Sl} \#\Sl
r_{\vl'}(i)\prod_{j=1}^d p_j = d\sum_{i\in \Sl}
r_{\vl'}(i)\prod_{j=1}^d p_j,
\]
and outside $\Pnn$, $\prod_{j=1}^d p_j=0$ so that the
equality still holds. Therefore $t_1 = u_1$.

Now $p_j\leq N+1$ so that $\prod_{j=1}^d p_j \leq (N+1)^d$ and
$|r_{\vl'}(i)|\leq 2 (N+1)^{d-1} $. Moreover, as $1\leq \la_i\leq N+1$ and $\la_d$ is known if the other $\la_i$
are known,  the number of elements $\vl'$ in
$\Pnn$ satisfies $\#\Pnn \leq (N+1)^{d-1}$. Thus the
numerator of $t_1$ and $u_1$ is $O(N^{3d-2})$ and that of $t_2$ and $u_2$ is
$O(N^{3d-3})$. To end the proof of the lemma, it is then sufficient to show
that $C_u = \Theta(N^{3d-1})$.

Let us write $N+1 = a(1+d(d+1))/2  + b$ with $a$ and $b$ natural integers and
$b < (1+d(d+1)) $. We then select $h_i$ for $i=1$ to $d$ such that $\sum h_i
= a/2$. The number of ways of partitioning $a/2$ in $d$ parts is ${a/2+d-1  \choose
d-1}$, and this is $\Theta(a^{d-1})= \Theta(N^{d-1})$. To each of these
partitions, we associate a different $\vl'$ in $\Pnn$ through $ \la_i = (d-i+1)
a+ \delta_{i=1} b + h_i$. For each of these $\vl'$, we have
$p_j=\la_j-\la_{j+1}\geq a/2$, so that $\prod_{j=1}^d p_j^2 = \Theta(N^{2d})$.
We may lower bound $C_u$ by the sum over these $\vl'$ of $\prod_{j=1}^d p_j^2$,
so that we have proved $C_u = \Theta(N^{3d-1})$. 
\end{proof}

\section{Evaluation of the constant in the speed of convergence and final
result}
\label{constante}

  The strategy we study is asymptotically optimal up
to a constant,  but a better constant can probably be obtained. Anything like
$c(\vl)=(\prod p_j)^{\alpha}$ with $\alpha\geq 1/2$ should yield the same rate,
though it would be more cumbersome to prove. Polynomials in the $p_j$ could
also bring some improvement. All the same we give in this section a quick evaluation of the
constant, that may serve as a benchmark for more precise strategies.

Write $p_j = (N+1) x_j$. Then, recalling our notation \ref{notation}, 
\begin{align*}
\prod_{j=1}^d p_j^2 & = (N+1)^{2d} \prod_{j=1}^d x_j^2 \\
r_{\vl'}(i) & = (N+1)^{d-1} \left( - x_{\{i\}}
+\delta_{i>1}x_{\{i-1\}} + O(N^{-1})\right).
\end{align*}
Similarly, the set of allowed $\vec{x}=(x_1,\dots,x_n)$ may be described as
\[\Sn=\left\{\vec{x} \,| \,x_j(N+1)\in\mathbb{N}; \sum_{j=1}^d
(d-j+1)x_j=1\right\}.\]

We may then rewrite:
\begin{align*}
u_2 & =\frac{ \sum_{\vec{x}\in\Sn} d \sum_{i=1}^d\left(  x_{\{i\}}
-\delta_{i>1}x_{\{i-1\}}  \right)^2} {d^2 (N+1)^{2}
\sum_{\vec{x}\in\Sn}\prod_{j=1}^dx_j^2}+O(N^{-3}) 
\\
t_2 &=
\frac{\sum_{\vec{x}\in\Sn}\left(x_{\{i\}}
- \delta_{i>1} x_{\{i-1\}} \right)^2}
{d^2 (N+1)^{2}
\sum_{\vec{x}\in\Sn}\prod_{j=1}^dx_j^2}+ O(N^{-3}).
\end{align*}
Subtracting, we obtain (the first sums being on $\Sn$)
\begin{align}
\label{interm}
& u_2-t_2 + O(N^{-3}) = \\
& 
\frac{\sum_{\vec{x}} 2d\left(\sum_{i=1}^{d}
(x_{\{i\}})^2- \sum_{i=2}^d x_{\{i\}}x_{\{i-1\}}
\right)-(d+1)(x_{\{d\}})^2}
{n^2\:d^2 
\sum_{\vec{x}}\prod_{j=1}^dx_j^2}.
\end{align} 

Now $\Sn$ is the intersection $\mathcal{S}$ of the lattice in $[0,1]^d$ with mesh
size $1/(N+1)$
with the hyperplane given by the equation $\sum (d-j+1)x_j =1 $. Therefore
the points of $\Sn$ are a regular paving of a flat $(d-1)$-dimensional volume,
with more and more points (we know that $\#\Sn= O(N^{d-1})$). Therefore both
denominator and numerator of  (\ref{interm}) are Riemannian sums with
respect to the Lebesgue measure, with a multiplicative constant that is the
same for both. Therefore we have proved:

\begin{thm}
\label{conclusion}
The estimator $\hat{U}$ corresponding to (\ref{weight}) has the following
risk:
\[
R_B(\hat{U}) = R_P(\hat{U}) = \mathbb{E}_{{\bf 1}_{\Cd}}\left[\Delta({\bf
1}_{\Cd},\hat{U})\right]= C N^{-2} + O(N^{-3})
\]
where $C$ is the fraction
\[
\frac{\int_{\mathcal{S}} 2d\left(\sum_{i=1}^{d} (x_{\{i\}})^2 -
\sum_{i=2}^d x_{\{i\}}x_{\{i-1\}}
\right)-(d+1) (x_{\{d\}})^2 \dd
\vec{x}}
{d^2 
\int_{\mathcal{S}}\prod_{j=1}^d x_j^2 \dd \vec{x}}.
\]
Up to a multiplicative constant, this risk is asymptotically optimal, both for
a Bayes uniform prior and for global pointwise estimation. 
\end{thm}

Numerical estimation, up to two digits, for the low dimensions yields:
\begin{align*}
 10 & \mathrm{~ for ~ } d=2 \\
 75 & \mathrm{~ for ~ } d=3 \\
 2.7\times10^2 & \mathrm{~ for ~ } d=4. 
\end{align*} 

\section{Conclusion}

We have given a strategy for estimating an unknown unitary
channel $U\in SU(d)$, and proved that the convergence rate of this strategy is
$1/N^2$. We have further proved that this rate is optimal, even if the constant
may be improved.

The interest of this result lies in that such rates are much faster than the
$1/N$ achieved in classical estimation and, though they had already been
obtained for
$SU(2)$, they were never before shown to hold for general $SU(d)$.

\begin{acknowledgments}
We are indebted to Manuel Ballester for an introduction to this question and for kindly providing the figure, and
to M\u ad\u alin Gu\c t\u a for numerous suggestions and extensive rereading.
\end{acknowledgments}


\end{document}